\documentclass[conference,twocolumn,final,letterpaper]{IEEEtran}%
\usepackage{amsfonts}
\usepackage{amsmath}
\usepackage{amssymb}
\usepackage[dvips]{graphicx}
\usepackage{cite}%
\newtheorem{theorem}{Theorem}

\newtheorem{case}{Case}

\newtheorem{remark}[theorem]{Remark}

\begin{document}

\title{Opportunistic Communications in an Orthogonal Multiaccess Relay\ Channel }
\pubid{\ }
\specialpapernotice{\ }%

\author{\authorblockN{Lalitha Sankar}
\authorblockA{WINLAB, Dept. of ECE\\
Rutgers University\\
North Brunswick, NJ 08902\\
lalitha@winlab.rutgers.edu}
\and\authorblockN{Yingbin Liang, H. Vincent Poor}
\authorblockA{Dept. of Electrical Engineering\\
Princeton University\\
Princeton, NJ 08544\\
\{yingbinl,poor\}@princeton.edu}
\and\authorblockN{Narayan Mandayam}
\authorblockA{WINLAB, Dept. of ECE\\
Rutgers University\\
North Brunswick, NJ 08902\\
narayan@winlab.rutgers.edu}}%
%

\maketitle
%

\begin{abstract}%

\footnotetext{This research was supported in part by the US\ National\ Science
Foundation under Grants ANI-03-38807, CCR-04-29724 and CNS-06-25637.}The
problem of resource allocation is studied for a two-user fading
\textit{orthogonal} multiaccess relay channel (MARC) where both users
(sources) communicate with a destination in the presence of a relay. A
half-duplex relay is considered that transmits on a channel orthogonal to that
used by the sources. The instantaneous fading state between every
transmit-receive pair in this network is assumed to be known at both the
transmitter and receiver. Under an average power constraint at each source and
the relay, the sum-rate for the achievable strategy of decode-and-forward (DF)
is maximized over all power allocations (policies) at the sources and relay.
It is shown that the sum-rate maximizing policy exploits the multiuser fading
diversity to reveal the optimality of \textit{opportunistic} channel use by
each user. A geometric interpretation of the optimal power policy is also presented.%

\end{abstract}%

\section{Introduction}

The multiaccess relay channel (MARC) is a network in which several users
(source nodes) communicate with a single destination in the presence of a
relay \cite{cap_theorems:KvW01}. The MARC is a model for relay-based
cooperation in a multiuser network where the users have limited power and
processing capabilities or need tangible incentives to cooperate. We model a
MARC with a half-duplex relay as an \textit{orthogonal} MARC where the relay
transmits on a channel orthogonal to that used by the sources (see
\cite{cap_theorems:Liang_Veeravalli02,cap_theorems:SKM02a}). The coding
strategies developed for the relay channel \cite{cap_theorems:CEG01} extend
readily to the MARC \cite{cap_theorems:SKM02a}. For example, the strategy of
\cite[Theorem~1]{cap_theorems:CEG01}, now often called
\textit{decode-and-forward} (DF), has a relay that decodes user messages
before forwarding them to the destination
\cite{cap_theorems:KGG_IT,cap_theorems:SKM_journal01}. Similarly, the strategy
in \cite[Theorem~6]{cap_theorems:CEG01}, now often called
\textit{compress-and-forward} (CF), has the relay quantize its output symbols
and transmit the resulting quantized bits to the destination
\cite{cap_theorems:SKM02a}.

\bigskip

We study the problem of resource allocation in a two-user ergodic fading
orthogonal\ MARC employing DF under the assumption that the instantaneous
fading state between each transmit-receive pair in this network is known at
both the transmitter and receiver. Resource allocation for a single-user
ergodic fading orthogonal relay channel employing DF and subject to an average
power constraint at the source and relay is studied in
\cite{cap_theorems:Liang_Veeravalli02} (see also \cite{cap_theorems:HMZ01}).
The authors formulate the problem as a \textit{max-min} optimization. They
draw parallels with the classic minimax optimization in hypothesis testing to
show that the optimal resource allocation achieves one of three solutions
depending on the joint fading statistics. The orthogonal MARC studied here is
a multiaccess generalization of the orthogonal relay channel in
\cite{cap_theorems:Liang_Veeravalli02}; however, the optimal policies
developed in \cite{cap_theorems:Liang_Veeravalli02} do not extend readily to
maximize the sum-rate of the MARC. For a two-user MARC, we show that the
DF\ sum-rate belongs to one of five disjoint cases or lies on the boundary of
any two of them. Our results reveal two interesting observations: 1)
analogously to a classic fading multiaccess channel
\cite{cap_theorems:Knopp_Humblet,cap_theorems:TH01}, the sum-rate optimal
policy for each case exploits the multiuser fading diversity to
opportunistically schedule users; 2) however, these optimal policies are not
necessarily water-filling solutions. Finally, we present a geometric
interpretation for each case to highlight the effects of node topology in the
analysis of multi-terminal networks.

\bigskip

The paper is organized as follows. In Section \ref{Section 2}, we model the
orthogonal MARC with Gaussian noise and fading. In Section \ref{section 3} we
present the rate region and determine the power policies that maximize the DF
sum-rate. Finally, we conclude in Section \ref{section 4}.

\section{\label{Section 2}Channel Model and Preliminaries}

\begin{figure}
[ptb]
\begin{center}
\includegraphics[
height=1.6855in,
width=3.1168in
]%
{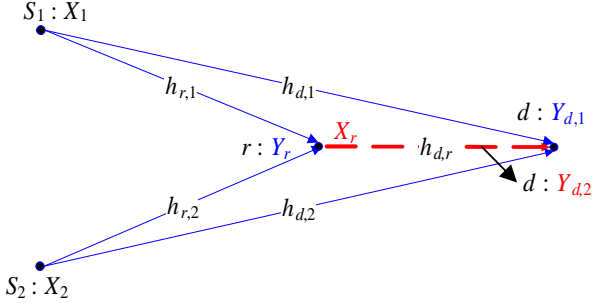}%
\caption{A two-user orthogonal MARC.}%
\label{Fig_1_model}%
\end{center}
\end{figure}

A two-user MARC consists of two source nodes numbered $1$ and $2$, a relay
node $r\,$,$\ $and a destination node $d$. We write $\mathcal{K}=\left\{
1,2\right\}  $ to denote the set of sources, $\mathcal{T}=\mathcal{K}%
\cup\left\{  r\right\}  $ to denote the set of transmitters, and
$\mathcal{D}=\left\{  r,d\right\}  $ to denote the set of receivers. In an
orthogonal MARC, the sources transmit to the relay and destination on one
channel, say channel 1, while the half-duplex relay transmits to the
destination on an orthogonal channel 2 as shown in Fig. \ref{Fig_1_model}. A
fraction $\theta$ of the total bandwidth resource is allocated to channel 1
while the remaining fraction $\overline{\theta}$ $=$ $1-\theta$ is allocated
to channel 2. In the fraction $\theta$, the source $k$ transmits the signal
$X_{k}$ while the relay and the destination receive $Y_{r}$ and $Y_{d,1}$
respectively. In the fraction $\overline{\theta}$, the relay transmits $X_{r}$
and the destination receives $Y_{d,2}$. In each time symbol (channel use), we
then have
\begin{align}
Y_{r}  &  =h_{r,1}X_{1}+h_{r,2}X_{2}+Z_{r}\label{GMARC_defn1}\\
Y_{d,1}  &  =h_{d,1}X_{1}+h_{d,2}X_{2}+Z_{d,1}\label{GMARC_defn2}\\
Y_{d,2}  &  =h_{d,r}X_{r}+Z_{d,2} \label{GMARC_defn3}%
\end{align}
where $Z_{r},$ $Z_{d,1},$ $Z_{d,2}$ are independent circularly symmetric
complex Gaussian noise random variables with zero means and unit variances. We
write \underline{$h$} to denote the vector of fading gains, $h_{k,m}$, for all
$k$ $\in$ $\mathcal{D}$ and $m$ $\in$ $\mathcal{T}$, $k$ $\not =$ $m$, such
that \underline{$h$} is a realization for a given channel use of a jointly
stationary and ergodic vector fading process \underline{$H$}. We assume that
the fraction $\theta$ is fixed \textit{a priori} and is known at all nodes.

\bigskip

Over $n$ uses of the channel, the source and relay tranmissions are
constrained in power according to%
\begin{equation}
\left.  \sum\limits_{i=1}^{n}\mathbb{E}(\left\vert X_{ki}\right\vert ^{2})\leq
n\overline{P}_{k}\right.  \text{ for all }k\in\mathcal{T}\text{.}
\label{GMARC_Pwr_def0}%
\end{equation}
Since the sources and relay know the fading states of the links on which they
transmit, they can allocate their transmitted signal power according to the
channel state information. We write $P_{k}(\underline{h})$ to denote the power
allocated as a function of the channel states $\underline{h}$ at the $k^{th}$
transmitter, for all $k\in\mathcal{T}$. For an ergodic fading channel,
(\ref{GMARC_Pwr_def0}) then simplifies to
\begin{equation}
\left.  \mathbb{E}(P_{k}(\underline{h}))\leq\overline{P}_{k}\right.  \text{
for all }k\in\mathcal{T} \label{GMARC_Pwr_defn}%
\end{equation}
where the expectation in (\ref{GMARC_Pwr_defn}) is over the joint distribution
$\underline{H}$. We write $\underline{P}\left(  \underline{h}\right)  $ to
denote a vector of power allocations with entries $P_{k}(\underline{h})$ for
all $k$ $\in$ ${\mathcal{T}}$, and define $\mathcal{P}$ to be the set of all
$\underline{P}\left(  \underline{h}\right)  $ whose entries satisfy
(\ref{GMARC_Pwr_defn}). For ease of notation, we henceforth omit the
functional dependence of $\underline{P}$ on $\underline{h}$. We use the
notation $C(x)$ $=$ $\log(1+x)$ where the logarithm is to the base 2, $\left(
x\right)  ^{+}=\max(x,0)$, and write $R_{\mathcal{S}}$ $=$ $%
{\textstyle\sum\nolimits_{k\in\mathcal{S}}}
R_{k}$ for any ${\mathcal{S}}$ $\subseteq$ ${\mathcal{K}}$.%

\section{\label{section 3}Sum-Rate Optimal Power Policy}

The DF\ rate region for a MARC with fixed channel gains and a full-duplex
relay is developed in \cite[Appendix A]{cap_theorems:KGG_IT} (see also
\cite{cap_theorems:SKM_journal01}). For a half-duplex MARC with a fixed
$\underline{h}$ and a fixed fraction $\theta$, the DF rate region includes an
additional conditioning on the half-duplex modes of the relay
\cite{cap_theorems:SKM02a} and is the set of all rate pairs $(R_{1},R_{2})$
that satisfy%
\begin{equation}
R_{k}\leq\min\left\{
\begin{array}
[c]{c}%
\theta C\left(  \frac{\left\vert h_{d,k}\right\vert ^{2}P_{k}}{\theta}\right)
+\overline{\theta}C\left(  \frac{\left\vert h_{d,r}\right\vert ^{2}P_{r}%
}{\overline{\theta}}\right)  ,\\
\theta C\left(  \frac{\left\vert h_{r,k}\right\vert ^{2}P_{k}}{\theta}\right)
\end{array}
\right\}  ,k=1,2 \label{GMARC_R1_fixh}%
\end{equation}
and
\begin{equation}
R_{1}+R_{2}\leq\min\left\{
\begin{array}
[c]{c}%
\theta C\left(  \sum\limits_{k=1}^{2}\frac{\left\vert h_{d,k}\right\vert
^{2}P_{k}}{\theta}\right)  +\overline{\theta}C\left(  \frac{\left\vert
h_{d,r}\right\vert ^{2}P_{r}}{\overline{\theta}}\right)  ,\\
\theta C\left(  \sum\limits_{k=1}^{2}\frac{\left\vert h_{r,k}\right\vert
^{2}P_{k}}{\theta}\right)
\end{array}
\right\}  . \label{GMARC_R12_fixh}%
\end{equation}
For a stationary and ergodic vector process $\underline{H}$, the channel in
(\ref{GMARC_defn1})-(\ref{GMARC_defn3}) can be modeled as a set of parallel
Gaussian orthogonal MARCs, one for each fading instantiation $\underline{h}$.
For a fixed $\underline{P}$, the DF rate bounds for this ergodic fading
channel are obtained by averaging the bounds in (\ref{GMARC_R1_fixh}) and
(\ref{GMARC_R12_fixh}) over all channel realizations. The DF rate region,
$\mathcal{R}_{DF}$, achieved over all $\underline{P}\in\mathcal{P}$, is given
by the following theorem.

\begin{theorem}
The DF rate region, $\mathcal{R}_{DF}$, achieved over an ergodic fading
orthogonal Gaussian MARC is
\begin{equation}
\mathcal{R}_{DF}=\bigcup\limits_{\underline{P}\in\mathcal{P}}\left\{
\mathcal{R}_{r}\left(  \underline{P}\right)  \cap\mathcal{R}_{d}\left(
\underline{P}\right)  \right\}  \label{GMARC_R_DF}%
\end{equation}
where, for all $\mathcal{S}\subseteq\mathcal{K}$, we have
\begin{equation}
\mathcal{R}_{r}\left(  \underline{P}\right)  =\left\{  \left(  R_{1}%
,R_{2}\right)  :R_{\mathcal{S}}\leq\theta\mathbb{E}C\left(  \frac
{\sum\limits_{k\in\mathcal{S}}\left\vert h_{r,k}\right\vert ^{2}P_{k}}{\theta
}\right)  \right\}  \label{DF_Rates_rel}%
\end{equation}
and%
\begin{equation}
\mathcal{R}_{d}\left(  \underline{P}\right)  =\left\{
\begin{array}
[c]{c}%
\left(  R_{1},R_{2}\right)  :R_{\mathcal{S}}\leq\theta\mathbb{E}C\left(
\frac{\sum\limits_{k\in\mathcal{S}}\left\vert h_{d,k}\right\vert ^{2}P_{k}%
}{\theta}\right) \\
+\overline{\theta}\mathbb{E}C\left(  \frac{\left\vert h_{d,r}\right\vert
^{2}P_{r}}{\overline{\theta}}\right)
\end{array}
\right\}  \label{DF_Rates_dest}%
\end{equation}

\end{theorem}

\begin{remark}
\label{Remark}The rate region $\mathcal{R}_{DF}$ is convex. This follows from
the convexity of the set $\mathcal{P}$ and the concavity of the $\log$ function.
\end{remark}

\bigskip

\begin{figure}
[ptb]
\begin{center}
\includegraphics[
height=1.9303in,
width=3.0761in
]%
{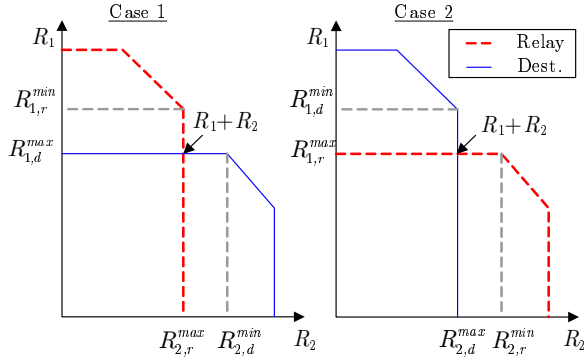}%
\caption{Rate region and sum-rate for case 1 and case 2.}%
\label{Fig_Case12}%
\end{center}
\end{figure}

\begin{figure}
[ptb]
\begin{center}
\includegraphics[
height=1.9242in,
width=3.4722in
]%
{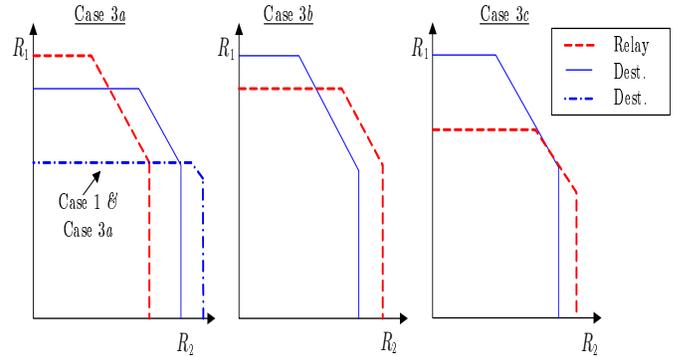}%
\caption{Rate region and sum-rate for cases $3a$, $3b$, and $3c$.}%
\label{Fig_Case3abc}%
\end{center}
\end{figure}

The region $\mathcal{R}_{DF}$ in (\ref{GMARC_R_DF}) is a union of the
intersections of the regions $\mathcal{R}_{r}(\underline{P})$ and
$\mathcal{R}_{d}(\underline{P})$ achieved at the relay and destination
respectively, where the union is over all $\underline{P}$ $\in$ $\mathcal{P}$.
Since $\mathcal{R}_{DF}$ is convex, each point on the boundary of
$\mathcal{R}_{DF}$ is obtained by maximizing the weighted sum $\mu_{1}R_{1}$
$+$ $\mu_{2}R_{2}$ over all $\underline{P}\in\mathcal{P}$, and for all
$\mu_{1}>0$, $\mu_{2}>0$. Specifically, we determine the optimal policy
$\underline{P}^{\ast}$ that maximizes the sum-rate $R_{1}+R_{2}$ when $\mu
_{1}$ $=$ $\mu_{2}$ $=$ $1$. Observe from (\ref{GMARC_R_DF}) that every point
on the boundary of $\mathcal{R}_{DF}$ results from the intersection of
$\mathcal{R}_{r}(\underline{P})$ and $\mathcal{R}_{d}(\underline{P})$ for some
$\underline{P}$. In Figs. \ref{Fig_Case12} and \ref{Fig_Case3abc} we
illustrate the five possible choices for the sum-rate resulting from such an
intersection. Case $1$ and case $2$ result when no rate pair on the sum-rate
plane achieved at one receiver lies within or on the boundary of the rate
region achieved at the other receiver (see Fig. \ref{Fig_Case12}). On the
other hand, cases $3a$, $3b$, and $3c$ result when there is more than one such
rate pair as shown in Fig. \ref{Fig_Case3abc}. Observe that case $3c$
corresponds to a boundary case where the sum-rate planes overlap. We also
consider six boundary cases where there is exactly one such rate pair that
serves as a transition between case $1$ or $2$ and one of cases $3a$, $3b\,$,
or $3c$. An example of a boundary case for case $1$ and case $3a$ is shown in
Fig. \ref{Fig_Case3abc}. We write $\mathcal{B}_{m}\subseteq\mathcal{P}$ to
denote the set of $\underline{P}$ that achieve case $i$, $i=1,2,3a,3b,3c$ and
$\mathcal{B}_{l,n}$, $l=1,2$, $n=3a,3b,3c$ to denote the set of $\underline
{P}$ satisfying each boundary case. We show in the sequel that the
optimization is simplified considerably when the conditions for each case are
defined such that the sets $\mathcal{B}_{i}$ and $\mathcal{B}_{l,n}$ are
disjoint for all $i,l,n$, and thus, are either open or half-open sets such
that no two sets share a boundary. Finally, we observe that cases $1$ and $2$
do not share a boundary since such a transition (see Fig. \ref{Fig_Case12})
requires passing through case $3a$ or $3b$ or $3c$.%

\bigskip

To determine the optimal $\underline{P}^{\ast}$, we first define the
conditions for each case and determine the policy $\underline{P}^{\left(
i\right)  }$ or $\underline{P}^{\left(  l,n\right)  }$ maximizing the sum-rate
for case $i$ or the boundary case $\left(  l,n\right)  $. We collect the six
boundary cases as the last case. The optimal policy for each case is
determined using Lagrange multipliers and the \textit{Karush}-\textit{Kuhn}%
-\textit{Tucker} (KKT) conditions \cite[5.5.3]{cap_theorems:BVbook01}.

\begin{case}
\label{Case_1}This case occurs when the power policy $\underline{P}%
\in\mathcal{B}_{1}$ achieves the relay and destination regions shown in Fig
\ref{Fig_Case12}. The maximum sum-rate achieved in this case is
\begin{equation}
\max_{\underline{P}\in\mathcal{B}_{1}}\left(  R_{1,d}^{\max}\left(
\underline{P}\right)  +R_{2,r}^{\max}\left(  \underline{P}\right)  \right)
\label{Case1_sumrate}%
\end{equation}
where $R_{k,m}^{\max}$ is the maximum rate achieved by user $k$ at receiver
$m$ $\in$ $\mathcal{D}$ in (\ref{DF_Rates_rel}) and (\ref{DF_Rates_dest}). The
open set $\mathcal{B}_{1}$ contains all $\underline{P}$ that satisfy%
\begin{equation}
R_{1,d}^{\max}\left(  \underline{P}\right)  <R_{1,r}^{\min}\left(
\underline{P}\right)  \text{ and }R_{2,r}^{\max}\left(  \underline{P}\right)
<R_{2,d}^{\min}\left(  \underline{P}\right)  \label{Case1_Conds}%
\end{equation}
where $R_{k,m}^{\min}$ is the rate achieved by user $k$ when it is the first
user to be successively decoded at a sum-rate corner point achieved at
receiver $m$. Since $\mathcal{B}_{1}$ is not known a priori, we determine the
optimal $\underline{P}^{\left(  1\right)  }$ maximizing $R_{1}+R_{2}$ in
(\ref{Case1_sumrate}) over $\mathcal{P}$. Expanding (\ref{Case1_sumrate})
using (\ref{DF_Rates_rel}) and (\ref{DF_Rates_dest}) and applying the Lagrange
multiplier rule and the KKT conditions, we obtain $P_{k}^{\left(  1\right)  }$
and $P_{r}^{\left(  1\right)  }$ as%
\begin{equation}%
\begin{array}
[c]{ll}%
P_{k}^{\left(  1\right)  }=\left(  \frac{\theta}{\nu_{k}\ln2}-\frac{\theta
}{\left\vert h_{m,k}\right\vert ^{2}}\right)  ^{+} & (k,m)=(1,d),(2,r)
\end{array}
\label{Case1_OptPolicy}%
\end{equation}
and
\begin{equation}
\text{ }P_{r}^{\left(  1\right)  }=\left(  \frac{\overline{\theta}}{\nu_{r}%
\ln2}-\frac{\overline{\theta}}{\left\vert h_{d,r}\right\vert ^{2}}\right)
^{+}\text{ } \label{Case1_OptPr}%
\end{equation}
where the water-filling level $\nu_{k}$, $k=1,2,r,$ is determined from
(\ref{GMARC_Pwr_defn}). To ensure that case \ref{Case_1} occurs, we require
that \underline{$P$}$^{\left(  1\right)  }$ $\in$ $\mathcal{B}_{1}$, i.e.,
(\ref{Case1_OptPolicy}) and (\ref{Case1_OptPr}) satisfy (\ref{Case1_Conds}).
Then, the concavity of the rate functions in (\ref{DF_Rates_rel}) and
(\ref{DF_Rates_dest}) suffices to show that \underline{$P$}$^{\left(
1\right)  }$ in (\ref{Case1_OptPolicy}) and (\ref{Case1_OptPr}) maximizes
(\ref{Case1_sumrate}). On the other hand, when \underline{$P$}$^{\left(
1\right)  }$ $\not \in $ $\mathcal{B}_{1}$, we show that $R_{1}+R_{2}$
achieves its maximum outside $\mathcal{B}_{1}$. Note that the expression for
$R_{1}+R_{2}$ for the other cases is not the same as that in
(\ref{Case1_sumrate}). The proof follows from the fact that $R_{1}+R_{2}$ in
(\ref{Case1_sumrate}) is a concave function of \underline{$P$} for all
$\underline{P}$ $\in$ $\mathcal{P}$. Thus, when \underline{$P$}$^{\left(
1\right)  }$ $\not \in $ $\mathcal{B}_{1}$, for every \underline{$P$} $\in$
$\mathcal{B}_{1}$ there exists a \underline{$P$}$^{^{\prime}}$ $\in$
$\mathcal{B}_{1}$ with a larger sum-rate. Combining this with the fact that
the sum-rate expressions are continuous while transitioning from one case to
another at the boundary of the open set $\mathcal{B}_{1}$, ensures that the
maximum sum-rate is achieved by some $\underline{P}$ $\not \in $
$\mathcal{B}_{1}$. We remark that similar arguments also apply to the
remaining cases, and will be omitted for brevity. Finally, we remark that
(\ref{Case1_Conds}) models a network geometry in which the destination and
source 1 are physically proximal, i.e., they form a cluster, and the relay and
source 2 form another cluster.
\end{case}

%

\begin{case}
The maximum sum-rate achieved for this case is
\begin{equation}
\max_{\underline{P}\in\mathcal{B}_{2}}\left(  R_{1,r}^{\max}\left(
\underline{P}\right)  +R_{2,d}^{\max}\left(  \underline{P}\right)  \right)
\label{Case2_sumrate}%
\end{equation}
where $\mathcal{B}_{2}$ contains all $\underline{P}$ that satisfy%
\begin{equation}
R_{1,r}^{\max}\left(  \underline{P}\right)  <R_{1,d}^{\min}\left(
\underline{P}\right)  \text{ and }R_{2,d}^{\max}\left(  \underline{P}\right)
<R_{2,r}^{\min}\left(  \underline{P}\right)  . \label{Case2_Conds}%
\end{equation}
As in case \ref{Case_1}, we can show that (\ref{Case2_sumrate}) is maximized
by setting the optimal $P_{k}^{(2)}$ and $P_{r}^{(2)}$ to the expressions in
(\ref{Case1_OptPolicy}) and (\ref{Case1_OptPr}) respectively with
$(k,m)=(1,r),(2,d)$ provided the resulting \underline{$P$}$^{\left(  2\right)
}$ satisfies (\ref{Case2_Conds}). Finally, we remark that the conditions in
(\ref{Case2_Conds}) model a network geometry in which the destination and
source 2 form one cluster while the relay and source 1 form another cluster.
\end{case}

\begin{case}
\label{Case_2}Consider the cases $3a,$ $3b,$ and $3c$ shown in Fig.
\ref{Fig_Case3abc}. The sum-rate optimization for all three cases simplifies
to
\begin{equation}
\max_{\underline{P}\in\mathcal{B}_{3}}\min\left(  \left(  R_{\mathcal{K}%
}\right)  _{r},\left(  R_{\mathcal{K}}\right)  _{d}\right)
\label{Case3_sumrate}%
\end{equation}
where $\left(  R_{\mathcal{K}}\right)  _{r}$ and $\left(  R_{\mathcal{K}%
}\right)  _{d}$ are the mutual information expressions in (\ref{DF_Rates_rel})
and (\ref{DF_Rates_dest}) respectively for $\mathcal{S}$ $=$ $\mathcal{K}$,
and $\mathcal{B}_{3}$ consists of $\underline{P}$ that do not satisfy
(\ref{Case1_Conds}) and (\ref{Case2_Conds}) either as strict inequalities or
with equality. We write $\mathcal{B}_{3}$ $=$ $\mathcal{B}_{3a}\cup
\mathcal{B}_{3b}\cup\mathcal{B}_{3c}$, where $\mathcal{B}_{i}$, $i=3a,3b,3c$
is defined for case $i$ below. The optimization in (\ref{Case3_sumrate}) is a
multiuser generalization of the single-user \textit{max-min} problem studied
in \cite{cap_theorems:Liang_Veeravalli02}. Extending the known general
solution to (\ref{Case3_sumrate}), the optimal policy $\underline{P}^{\left(
3\right)  }$ for this case satisfies one of following three conditions%
\begin{align}
&
\begin{array}
[c]{cc}%
\text{\textit{Case 3a}:} & \left(  R_{\mathcal{K}}\right)  _{r}|_{\underline
{P}^{\left(  3\right)  }}<\left(  R_{\mathcal{K}}\right)  _{d}|_{\underline
{P}^{\left(  3\right)  }}%
\end{array}
\label{Case3a_Cond}\\
&
\begin{array}
[c]{cc}%
\text{\textit{Case 3b}:} & \left(  R_{\mathcal{K}}\right)  _{r}|_{\underline
{P}^{\left(  3\right)  }}>\left(  R_{\mathcal{K}}\right)  _{d}|_{\underline
{P}^{\left(  3\right)  }}%
\end{array}
\label{Case3b_Cond}\\
&
\begin{array}
[c]{cc}%
\text{\textit{Case 3c}:} & \left(  R_{\mathcal{K}}\right)  _{r}|_{\underline
{P}^{\left(  3\right)  }}=\left(  R_{\mathcal{K}}\right)  _{d}|_{\underline
{P}^{\left(  3\right)  }}.
\end{array}
\label{Case3c_Cond}%
\end{align}
We proceed to study cases $3a$, $3b$, and $3c$ in detail. \newline\textit{Case
3a}: Maximizing $\left(  R_{\mathcal{K}}\right)  _{r}$ subject to
(\ref{GMARC_Pwr_defn}) results in the KKT conditions
\begin{equation}%
\begin{array}
[c]{cc}%
f_{k}\leq\nu_{k}\ln2, & k=1,2
\end{array}
\label{Case3a_KKT}%
\end{equation}
where
\begin{equation}
f_{k}=\left\vert h_{r,k}\right\vert ^{2}\left/  \left(  1+\sum\limits_{k=1}%
^{2}\frac{\left\vert h_{r,k}\right\vert ^{2}P_{k}}{\theta}\right)  \right.  .
\label{Case3a_fk}%
\end{equation}
For those channel states in which the two users do not see the same scaled
fading gains at the relay, i.e., $\left.  \left\vert h_{r,1}\right\vert
\right/  \nu_{1}$ $\not =$ $\left.  \left\vert h_{r,2}\right\vert \right/
\nu_{2}$, (\ref{Case3a_KKT}) reduces, for all $k,j\in\mathcal{K},$ $k$
$\not =$ $j$, and $i=3a$, to
\begin{equation}
P_{k}^{\left(  i\right)  }=\left\{
\begin{array}
[c]{cc}%
\left(  \frac{\theta}{\nu_{k}\ln2}-\frac{\theta}{\left\vert h_{r,k}\right\vert
^{2}}\right)  ^{+} & f_{k}>f_{j}\frac{\nu_{k}}{\nu_{j}},j\not =k\\
0 & \text{o.w.}%
\end{array}
\right.  . \label{Case3a_optP}%
\end{equation}
Observe that the optimal $P_{k}^{\left(  3a\right)  }$ in (\ref{Case3a_optP})
is an opportunistic water-filling solution that exploits the fading diversity
in a multiaccess channel with the relay as the receiver. The optimal policy at
the relay $P_{r}^{(3a)}$ is given by (\ref{Case1_OptPr}). For \underline{$P$%
}$^{\left(  3a\right)  }\in\mathcal{B}_{3}$, the requirement of satisfying
(\ref{Case3a_Cond}), i.e., \underline{$P$}$^{\left(  3a\right)  }$ $\in$
$\mathcal{B}_{3a}$, simplifies to a threshold condition $\overline{P}_{r}$ $>$
$P_{u}\left(  \overline{P}_{1},\overline{P}_{2}\right)  $ where $\overline
{P}_{k}$, $k\in\mathcal{T}$, is defined in (\ref{GMARC_Pwr_defn}) and the
threshold $P_{u}\left(  \overline{P}_{1},\overline{P}_{2}\right)  $ is
obtained by setting (\ref{Case3a_Cond}) to an equality. When \underline{$P$%
}$^{\left(  3a\right)  }$ $\in$ $\mathcal{B}_{3}$ but \underline{$P$%
}$^{\left(  3a\right)  }$ $\not \in $ $\mathcal{B}_{3a}$, $R_{1}+R_{2}$ is
maximized by either \textit{case 3b} or \textit{case 3c}. For $\underline
{P}^{\left(  3a\right)  }$ $\not \in $ $\mathcal{B}_{3}$, as argued in case
\ref{Case_1}, the sum-rate is not maximized by any $\underline{P}$ $\in$
$\mathcal{B}_{3}$. Finally, the condition in (\ref{Case3a_Cond}) suggests a
geometry where the sources and destination are clustered. \newline\textit{Case
3b} : The KKT conditions and optimal $P_{k}^{\left(  3b\right)  }$ for this
case maximize $\left(  R_{\mathcal{K}}\right)  _{d}$ and are given by
(\ref{Case3a_KKT}) and (\ref{Case3a_optP}) respectively with the subscript
`$r$' in (\ref{Case3a_fk}) changed to `$d$' for all $k$ and $i=3b$. Further,
(\ref{Case3a_KKT}) and (\ref{Case3a_fk}) also hold for the relay node, $k=r$,
and simplifies to the water-filling solution in (\ref{Case1_OptPr}). Thus, as
in case $3a$, it is optimal to time-duplex the users except that this is now
based on their scaled fading gains to the destination. For \underline{$P$%
}$^{\left(  3b\right)  }$ $\in$ $\mathcal{B}_{3}$, satisfying
(\ref{Case3b_Cond}), i.e., \underline{$P$}$^{\left(  3b\right)  }$ $\in$
$\mathcal{B}_{3b}$, reduces to satisfying the threshold condition
$\overline{P}_{r}$ $<$ $P_{l}\left(  \overline{P}_{1},\overline{P}_{2}\right)
$ where $P_{l}\left(  \overline{P}_{1},\overline{P}_{2}\right)  $ is
determined by setting (\ref{Case3b_Cond}) to an equality. Finally,
(\ref{Case3b_Cond}) implies a geometry in which the sources are clustered
closer to the relay than to the destination. \newline\textit{Case 3c
}(\textit{equalizer policy}): Maximizing $\left(  R_{\mathcal{K}}\right)
_{r}$ over all \underline{$P$}, subject to (\ref{Case3c_Cond}) gives
\underline{$P$}$^{\left(  3c\right)  }$. This case occurs when \underline{$P$%
}$^{\left(  3c\right)  }$ $\in$ $\mathcal{B}_{3c}$, i.e. $P_{l}\left(
\overline{P}_{1},\overline{P}_{2}\right)  $ $\leq$ $\overline{P}_{r}$ $\leq$
$P_{u}\left(  \overline{P}_{1},\overline{P}_{2}\right)  $. The resulting KKT
conditions take the form in (\ref{Case3a_KKT}) such that for $k=1,2$,%
\begin{equation}
f_{k}=\frac{\left(  1-\alpha\right)  \left\vert h_{r,k}\right\vert ^{2}%
}{1+\sum\limits_{j=1}^{2}\left.  \left\vert h_{r,j}\right\vert ^{2}%
P_{j}\right/  \theta}+\frac{\alpha\left\vert h_{d,k}\right\vert ^{2}}%
{1+\sum\limits_{j=1}^{2}\left.  \left\vert h_{d,j}\right\vert ^{2}%
P_{j}\right/  \theta}. \label{Case3c_fk}%
\end{equation}
For $k=r$, $f_{r}$ is obtained by replacing $h_{r,k}$ in (\ref{Case3a_fk}) by
$h_{d,r}$ and scaling by $\alpha$. The Lagrange multiplier $\alpha$ accounts
for the boundary condition in (\ref{Case3c_Cond}) and is computed by
evaluating (\ref{Case3c_Cond}) at \underline{$P$}$^{\left(  3c\right)  }$. The
relay's optimal policy simplifies to the water-filling solution in
(\ref{Case1_OptPr}) with the first term scaled by $\alpha$. For $\left\vert
h_{m,1}\right\vert /\nu_{1}$ $\not =$ $\left\vert h_{m,2}\right\vert /\nu_{2}%
$, $m=r$ or $d$, (\ref{Case3a_KKT}) simplifies, for $f_{k}$ in
(\ref{Case3c_fk}) and for all $k,j\in\mathcal{K},$ $k$ $\not =$ $j$, and
$i=3c$ as
\begin{equation}
P_{k}^{\left(  i\right)  }=\left\{
\begin{array}
[c]{cc}%
\left(  \text{root of }f_{k}|_{P_{j}=0}\right)  ^{+} & f_{k}>f_{j}\\
0 & \text{o.w.}%
\end{array}
\right.  \label{Case3c_OptP}%
\end{equation}
From (\ref{Case3c_OptP}), we see that user $k$, $k=1,2,$ transmits
opportunistically over those channel states where a function $f_{k}$ of its
fading gains and power is larger than that of the other user. Note that the
optimal policy $P_{k}^{\left(  3c\right)  }$ in (\ref{Case3c_OptP}) is no
longer a water-filling solution at the two sources. Finally, we remark that
this boundary case occurs for a range of geometries that transition from the
clustered geometry of case $3a$ to that of case $3b$.
\end{case}

\begin{case}
\textit{Boundary Cases}: We now consider the six boundary cases and summarize
the optimal \underline{$P$}$^{\left(  l,n\right)  }$, $l=1,2$, $n=3a$, $3b$,
$3c$, for each case. As with case $3c$, we observe that the geometries for
these boundary cases also straddle the clustered geometries of the two cases
involved. \newline\textit{Case }$1$ and \textit{Case }$3a$: The sum-rate
optimization for this case simplifies to%
\begin{equation}
\max_{P\in\mathcal{B}_{1,3a}}\left(  R_{\mathcal{K}}\right)  _{r}\text{ s.t.
}\left(  R_{\mathcal{K}}\right)  _{r}=R_{1,d}^{\max}+R_{2,r}^{\max}
\label{CB_1_3a}%
\end{equation}
where $\mathcal{B}_{1,3a}$ is the set of all \underline{$P$} that satisfy
(\ref{Case3a_Cond}), (\ref{CB_1_3a}), and one of the inequalities in
(\ref{Case1_Conds}) (see Fig. \ref{Fig_Case3abc}). Note that from
(\ref{DF_Rates_rel}), we can write $\left(  R_{\mathcal{K}}\right)
_{r}=R_{1,r}^{\min}+R_{2,r}^{\max}$ thus simplifying the condition in
(\ref{CB_1_3a}) to $R_{1,r}^{\min}=R_{1,d}^{\max}$ as shown in Fig.
\ref{Fig_Case3abc}. As before, we obtain the KKT conditions in
(\ref{Case3a_KKT}) for $k=1,2$, $(k,m)=(1,d),(2,r)$, and%
\begin{equation}
f_{k}=\frac{\left(  1-\alpha\right)  \left\vert h_{r,k}\right\vert ^{2}%
}{1+\sum\limits_{j=1}^{2}\frac{\left\vert h_{r,j}\right\vert ^{2}P_{j}}%
{\theta}}+\frac{\alpha\left\vert h_{m,k}\right\vert ^{2}}{1+\frac{\left\vert
h_{m,k}\right\vert ^{2}P_{k}}{\theta}} \label{CB_1_3a_KKTs}%
\end{equation}
where $\alpha$ is the Lagrange multiplier satisfying the boundary condition in
(\ref{CB_1_3a}). When $\left\vert h_{m,1}\right\vert /\nu_{1}$ $\not =$
$\left\vert h_{m,2}\right\vert /\nu_{2}$, $m=r$ or $d$, (\ref{Case3a_KKT})
simplifies $P_{k}^{\left(  1,3a\right)  }$ to the opportunistic
non-water-filling solution in (\ref{Case3c_OptP}) with $f_{k}$ in
(\ref{CB_1_3a_KKTs}), and for all $k,j\in\mathcal{K},$ $k$ $\not =$ $j$,
$(k,m)$ $=$ $(1,d),(2,r)$, $(l,n)$ $=$ $(1,3a)$. The optimal $P_{r}^{\left(
1,3a\right)  }$ is given by (\ref{Case1_OptPr}) with the first term scaled by
$\alpha$. \newline\textit{Case }$2$ and \textit{Case }$3a$: The sum-rate
optimization for this case is%
\begin{equation}
\max_{P\in\mathcal{B}_{2,3a}}\left(  R_{\mathcal{K}}\right)  _{r}\text{ s.t.
}R_{2,r}^{\min}=R_{2,d}^{\max} \label{CB_2_3a}%
\end{equation}
where $\mathcal{B}_{2,3a}$ is the set of all \underline{$P$} that satisfy
(\ref{Case3a_Cond}), (\ref{CB_2_3a}), and the remaining inequality in
(\ref{Case2_Conds}). The resulting KKT conditions in (\ref{Case3a_KKT}) use
$f_{k}$, $k=1,2,$ defined in (\ref{CB_1_3a_KKTs}) but with $(k,m)=(1,r),(2,d)$%
. Note that $\alpha$ captures the equality condition in (\ref{CB_2_3a}). For
$\left\vert h_{m,1}\right\vert /\nu_{1}$ $\not =$ $\left\vert h_{m,2}%
\right\vert /\nu_{2}$, $m=r$ or $d$, and $(k,m)=(1,r),(2,d)$, $P_{k}^{\left(
2,3a\right)  }$ at each source is given by the opportunistic policy in
(\ref{Case3c_OptP}). The optimal relay policy $P_{r}^{\left(  2,3a\right)  }$
is the same as that obtained in $case$ $1$ and $case$ $3a$. \newline%
\textit{Case }$1$ and \textit{Case }$3b$: The sum-rate optimization for this
case is
\begin{equation}
\max_{P\in\mathcal{B}_{1,3b}}\left(  R_{\mathcal{K}}\right)  _{d}\text{ s.t.
}R_{2,d}^{\min}=R_{2,r}^{\max} \label{CB_1_3b}%
\end{equation}
where $\mathcal{B}_{1,3b}$ is the set of all \underline{$P$} that satisfy
(\ref{Case3b_Cond}), (\ref{CB_1_3b}), and the remaining inequality in
(\ref{Case1_Conds}). The resulting KKT conditions satisfy (\ref{Case3a_KKT})
where $f_{k}$, $k=1,2,$ is given by (\ref{CB_1_3a_KKTs}) with the subscript
`$r$' replaced by `$d$'. Note that $\alpha$ captures the boundary condition in
(\ref{CB_1_3b}). For $\left\vert h_{m,1}\right\vert /\nu_{1}$ $\not =$
$\left\vert h_{m,2}\right\vert /\nu_{2}$, $m=r$ or $d$, and
$(k,m)=(2,r),(1,d)$, $P_{k}^{\left(  1,3b\right)  }$ at each source is given
by the opportunistic non-water-filling solution in (\ref{Case3c_OptP}). The
optimal relay policy $P_{r}^{\left(  1,3b\right)  }$ simplifies to the
water-filling solution in (\ref{Case1_OptPr}).\newline\textit{Case }$2$ and
\textit{Case }$3b$: The sum-rate optimization for this case simplifies to%
\begin{equation}
\max_{P\in\mathcal{B}_{2,3b}}\left(  R_{\mathcal{K}}\right)  _{d}\text{ s.t.
}R_{1,d}^{\min}=R_{1,r}^{\max} \label{CB_2_3b}%
\end{equation}
where $\mathcal{B}_{2,3b}$ is the set of all \underline{$P$} that satisfy
(\ref{Case3b_Cond}), (\ref{CB_2_3b}), and the remaining inequality in
(\ref{Case2_Conds}). We remark that the KKT conditions are the same as
(\ref{Case3a_KKT}) where $f_{k}$, $k=1,2,$ is given by (\ref{CB_1_3a_KKTs})
with `$r$' replaced by `$d$' and $(k,m)=(1,r),(2,d)$. The resulting
$P_{k}^{\left(  2,3b\right)  }$, $k=1,2$ are given by the opportunistic
policies in (\ref{Case3c_OptP}). Finally, the optimal $P_{r}^{\left(
2,3b\right)  }$ is the same as that obtained in $case$ $1$ and $case$ $3b$.
\newline\textit{Case }$1$ and \textit{Case }$3c$: The sum-rate optimization
for this case is%
\begin{equation}
\max_{P\in\mathcal{B}_{1,3c}}\left(  R_{\mathcal{K}}\right)  _{r}\text{ s.t.
}R_{2,d}^{\min}=R_{2,r}^{\max}\text{ and }R_{1,d}^{\max}=R_{1,r}^{\min}\text{
} \label{CB_1_3c}%
\end{equation}
where $\mathcal{B}_{1,3c}$ is the set of all \underline{$P$} that satisfy
(\ref{Case3c_Cond}), (\ref{CB_1_3c}), and the remaining inequality in
(\ref{Case1_Conds}). The resulting KKT conditions satisfy (\ref{Case3a_KKT})
for $k=1,2$, and $(k,m)=(1,d),(2,r)$ where
\begin{equation}
f_{k}=\frac{\alpha_{3}\left\vert h_{r,k}\right\vert ^{2}}{1+\sum
\limits_{j=1}^{2}\frac{\left\vert h_{r,j}\right\vert ^{2}P_{j}}{\theta}}%
+\frac{\alpha_{2}\left\vert h_{d,k}\right\vert ^{2}}{1+\sum\limits_{j=1}%
^{2}\frac{\left\vert h_{d,j}\right\vert ^{2}P_{j}}{\theta}}+\frac{\alpha
_{1}\left\vert h_{m,k}\right\vert ^{2}}{1+\frac{\left\vert h_{m,k}\right\vert
^{2}P_{k}}{\theta}}%
\end{equation}
and $\alpha_{1}$, $\alpha_{2}$, and $\alpha_{3}$ $=$ $1$ $-$ $\alpha_{1}$ $-$
$\alpha_{2}$ are Lagrange multipliers that capture the boundary conditions in
(\ref{CB_1_3c}). When $\left\vert h_{m,1}\right\vert /\nu_{1}$ $\not =$
$\left\vert h_{m,2}\right\vert /\nu_{2}$, $m=r$ or $d$, for all $k,j\in
\mathcal{K},$ $k$ $\not =$ $j$, and pairs $(k,m)$ $=$ $(1,d),(2,r)$, the
optimal $P_{k}^{\left(  1,3c\right)  }$, $k=1,2$, is given by the
opportunistic policy in (\ref{Case3c_OptP}) while $P_{r}^{\left(  1,3c\right)
}$ is given by the water-filling solution in (\ref{Case1_OptPr}) with the
first term scaled by $(\alpha_{1}+\alpha_{2})$. \newline\textit{Case }$2$ and
\textit{Case }$3c$: The sum-rate optimization for this case is%
\begin{equation}
\max_{P\in\mathcal{B}_{2,3c}}\left(  R_{\mathcal{K}}\right)  _{r}\text{ s.t.
}R_{2,d}^{\max}=R_{2,r}^{\min}\text{ and }R_{1,d}^{\min}=R_{1,r}^{\max}\text{
} \label{CB_2_3c}%
\end{equation}
where $\mathcal{B}_{2,3c}$ is the set of all \underline{$P$} that satisfy
(\ref{Case3c_Cond}), (\ref{CB_2_3c}), and the remaining inequality in
(\ref{Case2_Conds}). The resulting KKT conditions and optimal policy
\underline{$P$}$^{(2,3c)}$ are the same as in \textit{case} $1$ and
\textit{case} $3c$ but with $(k,m)$ $=$ $(1,r),(2,d)$.
\end{case}

\bigskip

Finally, the optimal $\underline{P}^{\ast}$ is given by the following theorem.

\begin{theorem}
The $\underline{P}^{\ast}$ that maximizes the sum-rate is obtained by
computing $\underline{P}^{(m)}$ or $\underline{P}^{(j,k)}$ starting from case
$1$ and proceeding one case at a time, until for some case the corresponding
$\underline{P}^{(m)}$ or $\underline{P}^{(j,k)}$ satisfies the case conditions.
\end{theorem}

\section{\label{section 4}Summary and Future Work}

We have developed the power policy that maximizes the sum-rate of a two-user
orthogonal MARC. We have shown that the optimal policy is a function of the
channel statistics and network geometry and can be classified into two broad
categories. The first category involves cases where each user is clustered
with a different receiver as a result of which the sum-rate decouples into
independent terms for each user and the relay. The second category includes
the cases where the two users are clustered with one of the receivers as well
as the boundary cases. The first category admits the classic water-filling
solution at each user and the relay. The optimal policies for the second
category do not always result in a water-filling solution at the sources;
however, they reveal the optimality of exploiting the multiuser fading
diversity to opportunistically schedule users. Our results can be generalized
to a $K$-user orthogonal MARC with $K>2$. Finally, one could also consider the
resource allocation problem for CF where the challenge lies in solving a
non-convex optimization problem.

\bibliographystyle{IEEEtran}
\bibliography{MARC_refs}

\end{document}